%% file: main.tex
\pdfoutput=1

\documentclass[11pt]{article}

\usepackage[]{EMNLP2022}

\usepackage{times}
\usepackage{latexsym}

\usepackage[T1]{fontenc}

\usepackage[utf8]{inputenc}

\usepackage{microtype}

\usepackage{inconsolata}

\title{RACE: Retrieval-Augmented Commit Message Generation}

\newcommand\corrauthorfootnote[1]{%
  \begingroup
  \renewcommand\thefootnote{}\footnote{\textsuperscript{\S}#1}%
  \addtocounter{footnote}{-1}%
  \endgroup
}

\newcommand\notedauthorfootnote[1]{%
  \begingroup
  \renewcommand\thefootnote{}\footnote{\textsuperscript{\dag}#1}%
  \addtocounter{footnote}{-1}%
  \endgroup
}

\author{
Ensheng Shi\textsuperscript{a}
Yanlin Wang\textsuperscript{b,\S,\dag}
Wei Tao\textsuperscript{c}
Lun Du\textsuperscript{d}
\\
\textbf{Hongyu Zhang\textsuperscript{e}}\
\textbf{Shi Han\textsuperscript{d}}
\textbf{Dongmei Zhang\textsuperscript{d}}
\textbf{Hongbin Sun\textsuperscript{a,\S}}
\\
\textsuperscript{a}Xi'an Jiaotong University \quad
\textsuperscript{b}School of Software Engineering, Sun Yat-sen University   \\
\textsuperscript{c}Fudan University \quad
\textsuperscript{d}Microsoft Research \quad
\textsuperscript{e}The University of Newcastle
\\
{\tt s1530129650@stu.xjtu.edu.cn, hsun@mail.xjtu.edu.cn}\\
{\tt wangylin36@mail.sysu.edu.cn, wtao18@fudan.edu.cn}\\
{\tt \{lun.du, shihan, dongmeiz\}@microsoft.com}\\
{\tt hongyu.zhang@newcastle.edu.au}
}

\input{tool}
\usepackage{graphics}
\usepackage{arydshln}
\usepackage{listings}

\begin{document}
\maketitle

\begin{abstract}
Commit messages are important for software development and maintenance. Many neural network-based approaches have been proposed and shown promising results on automatic commit message generation. However, the generated commit messages could be repetitive or redundant. In this paper, we propose RACE, a new retrieval-augmented neural commit message generation method, which treats the retrieved similar commit as an exemplar and leverages it to generate an accurate commit message. As the retrieved commit message may not always accurately describe the content/intent of the current code diff, we also propose an \textit{exemplar guider}, which learns the semantic similarity between the retrieved and current code diff and then guides the generation of commit message based on the similarity. We conduct extensive experiments on a large public dataset with five programming languages. Experimental results show that RACE can outperform all baselines. Furthermore, RACE can boost the performance of existing Seq2Seq models in commit message generation. Our data and source code are available at \url{https://github.com/DeepSoftwareAnalytics/RACE}.
\end{abstract}

\section{Introduction}
In software development and maintenance\corrauthorfootnote{Yanlin Wang and Hongbin Sun are the corresponding authors.}, source code is frequently changed\notedauthorfootnote{Work done during the author’s employment at Microsoft Research Asia}. In practice, code changes are often documented as natural language commit messages, which summarize what (content) the code changes are or why (intent) the code is changed~\cite{BuseW10,Cortes-CoyVAP14}.
High-quality commit messages are essential to help developers understand the evolution of software without diving into implementation details, which can save a large amount of time and effort in software development and maintenance~\cite{DiasBGCD15, BarnettBBL15}. However, it is difficult to write high-quality commit messages due to lack of time, clear motivation, or experienced skills. Even for seasoned developers, it still poses a considerable amount of extra workload to write a concise and informative commit message for massive code changes~\cite{NieGZLLX21}. It is also reported that around 14\% of commit messages over 23,000 projects in SourceForge are left empty~\cite{0001NRN13}. Thus, automatically generating commit messages becomes an important task.

Over the years, many approaches have been proposed to automatically generate commit messages. Early studies~\cite{ShenSLYH16,Cortes-CoyVAP14} are mainly based on predefined rules or templates, which may not cover all situations or comprehensively infer the intentions behind code changes. Later, some studies~\cite{LiuXHLXW18,HuangZCXLL17,HuangJZCZT20} adopt information retrieval (IR) techniques to reuse commit messages of similar code changes. They can take advantage of similar examples, but the reused commit messages might not correctly describe the content/intent of the current code change. 
Recently, some Seq2Seq-based neural network  models~\cite{LoyolaMM17,JiangAM17,Xu00GT019,LiuLZFDQ19,jung2021commitbert} have been proposed to understand code diffs and generate the high-quality commit messages. These approaches show promising performance, but they tend to generate high-frequency and repetitive tokens and the generated commit messages have the problem of insufficient information and poor readability~\cite{WangXLHWG21,LiuXHLXW18}. 
Some studies~\cite{Liu20Atom,WangXLHWG21} also explore the combination of neural-based and IR-based techniques. \citet{Liu20Atom} propose an approach to rank the retrieved commit message (obtained by a simple IR-based model) and the generated commit message (obtained by a neural network model). \citet{WangXLHWG21} propose to use the similar code diff as auxiliary information in the inference stage,  while the model is not trained to learn how to effectively utilize the information of retrieval results. Therefore, both of them fail to take advantage of the information of retrieved similar results well.

In this paper, we propose a novel model 
\textbf{\Our} (\underline{R}etrieval-\underline{A}ugmented \underline{C}ommit m\underline{E}ssage generation), which retrieves a similar commit message as an exemplar, guides the neural model to learn the content of the code diff and the intent behind the code diff, and generates the readable and informative commit message. The key idea of our approach is retrieval and augmentation. Specifically, we first train a code diff encoder to learn the semantics of code diffs and encode the code diff into high-dimensional semantic space. Then, we retrieve the semantically similar code diff paired with the commit message on a large parallel corpus based on the similarity measured by vectors' distance. Next, we treat the similar commit message as an exemplar and leverage it to guide the neural-based models to generate an accurate commit message. However, the retrieved commit messages may not accurately describe the content/intent of current code diffs and may even contain wrong or irrelevant information. To avoid the retrieved samples dominating the processing of commit message generation, we propose an \textit{exemplar guider}, which first learns the semantic similarity between the retrieved and current code diff and then leverages the information of the exemplar based on the learned similarity to guide the commit message generation.

To evaluate the effectiveness of \Our{}, we conduct experiments on a large-scale dataset \mcmd~\cite{tao2021evaluation} with five programming language (\java, \csharp, \cpp, \python and \javascript) and compare \Our{} with 11 state-of-the-art approaches. 
Experimental results show that: (1) \Our{} significantly outperforms existing state-of-the-art approaches in terms of four metrics (BLUE, Meteor, Rouge-L and Cider) on the commit message generation. (2) \Our{} can boost the performance of existing Seq2Seq models in commit message generation. For example, it can improve the performance of \NMT~\cite{LoyolaMM17}, \commitbert~\cite{jung2021commitbert}, \codetf-small~\cite{0034WJH21} and \codetf-base~\cite{0034WJH21} by 43\%, 11\%, 15\%, and 16\% on average in terms of BLEU, respectively.
In addition, we also conduct human evaluation to confirm the effectiveness of \Our{}. 

We summarize the main contributions of this paper as follows:

\begin{itemize}
 
    \item We propose a retrieval-augmented neural commit message generation model, which treats the retrieved similar commit as an exemplar and leverages it to guide
    neural network model to generate informative and readable commit messages. 
    
    \item We apply our retrieval-augmented framework to four existing neural network-based approaches (\NMT, \commitbert, \codetf-small, and \codetf-base) and greatly boost their performance.
    
    \item We perform extensive experiments including human evaluation on a large multi-programming-language dataset and the results confirm the effectiveness of our approach over state-of-the-art approaches.
\end{itemize}

\section{Related Work}
\label{sec:related_work}
\revised{Code intelligence, which leverages machine learning especially deep learning-based method to understand source code, is an emerging topic and has obtained the promising results in many software engineering tasks, such as code summarization~\cite{zhangretrieval20,shi2021cast,shia2022evaluation,wang2020cocogum} and code search~\cite{GuZ018,du2021single,shi2022enhancing}.
Among them, commit message generation plays an important role in the software evolution.} 

In early work, information retrieval techniques are introduced to commit message generation~\cite{LiuXHLXW18,HuangZCXLL17,HuangJZCZT20}. For instance, 
ChangeDoc~\cite{HuangJZCZT20} retrieves the most similar commits according to the syntax or semantics in the code diff and reuses commit messages of similar code diffs. \textbf{\NNGen}~\cite{LiuXHLXW18} is a simple yet effective retrieval-based method using the nearest neighbor algorithm. It firstly recalls the top-k similar code diffs in the parallel corpus based on cosine similarity between bag-of-words vectors of code diffs. Then select the most similar result based on BLEU scores between each of them (top-k results) and the input code diff. These approaches can reuse similar examples and the reused commit messages are usually readable and understandable.

Recently, many neural-based approaches~\cite{LoyolaMM17,JiangAM17,Xu00GT019,LiuLZFDQ19,Liu20Atom,jung2021commitbert,dong2022fira,NieGZLLX21,WangXLHWG21} have been used to learn the semantic of code diffs and translate them into commit messages. For example, \textbf{\NMT{}}~\cite{LoyolaMM17} and \textbf{\Commitgen}~\cite{JiangAM17} treat the code diffs as plain texts and adopt the Seq2Seq neural network with different attention mechanisms to translate them into commit messages. \textbf{\Codisum}~\cite{Xu00GT019} extracts both code structure and code semantics from code diffs and jointly models them with a multi-layer bidirectional GRU to better learn the representations of code diffs. 
\textbf{\Ptrnet}~\cite{LiuLZFDQ19} incorporates the pointer-generator network into the Seq2Seq model to handle out-of-vocabulary (OOV) words. \textbf{\commitbert} leverage  CodeBERT~\cite{FengGTDFGS0LJZ20}, a pre-trained language model for source code, to learn the semantic representations of code diffs and adopt a Transformer-based~\cite{VaswaniSPUJGKP17} decoder to generate the commit message. These approaches show promising results on the generation of commit messages.

\revised{Recently, introducing retrieved relevant results into the training process has been found useful in most generation tasks~\cite{LewisPPPKGKLYR020,cikmYuXC21,wei2020retrieve}.} Some studies~\cite{Liu20Atom,WangXLHWG21} also explore the combination of neural-based models and IR-based techniques to generate commit messages. \textbf{\ATOM}~\cite{Liu20Atom} ensembles the neural-based model and the IR-based technique through the hybrid ranking. Specifically, it uses BiLSTM to encode ASTs paths extracted from ASTs of code diffs and adopt a decoder to generate commit messages. It also uses TF-IDF technique to represent code diffs as vectors and retrieves the most similar commit message based on cosine similarity. The generated and retrieved commit messages are finally prioritized by a hybrid ranking module.
\textbf{\Corec}~\cite{WangXLHWG21} is also a hybrid model and only considers the retrieved result during the inference. Specifically, at the training stage, they use an encoder-decoder neural model to encode the input code diffs by an encoder and generate commit messages by a decoder. At the inference stage, they first use the trained encoder to retrieve the most similar code diff from the training set. Then they reuse a trained encoder-decoder to encode the input and retrieved code diff, combine the probability distributions (obtained by two decoders) of each word, and generate the final commit message step by step.
In summary, \ATOM does not learn to refine the retrieved results or the generated results, and \Corec is not trained to utilize the information of retrieval results. Therefore, both of them fail to take full advantage of the retrieved similar results. In this paper, we treat the retrieved similar commit as an exemplar and train the model to leverage the exemplar to enhance commit message generation.

\begin{figure*}[ht]
\centering
\includegraphics[width=0.8\linewidth]{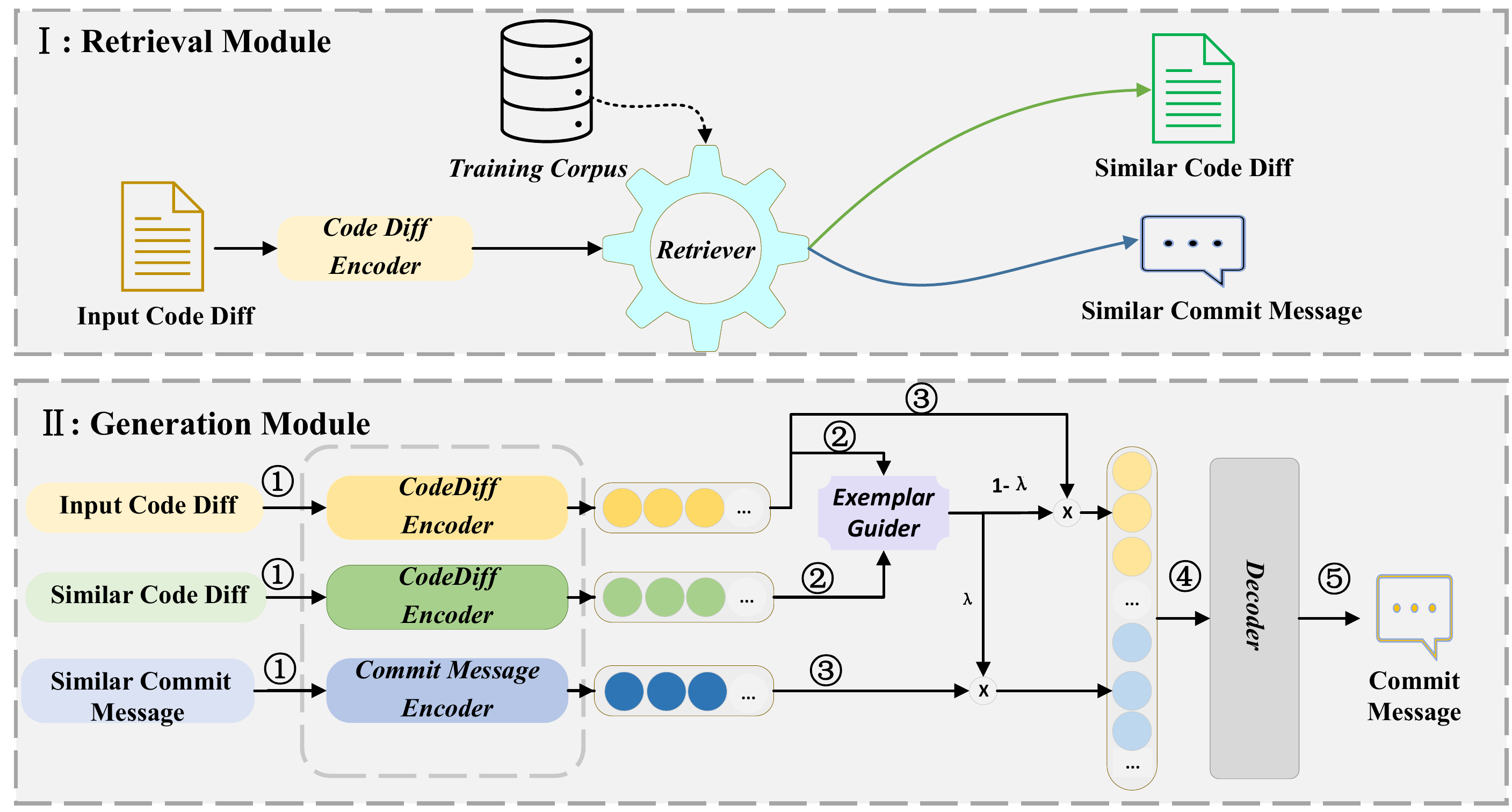}
\caption{The architecture of \Our{}. It includes two modules: retrieval module and generation module. The retrieval module is used to retrieve the most similar code diff and commit message. The generation module leverages the retrieved result to enhance the performance of neural network models. }
\label{fig:model}
\end{figure*}

\section{Proposed Approach}

The overview of \Our is shown in \Fig~\ref{fig:model}. It includes two modules: retrieval module and generation module. Specifically, \Our firstly retrieves the most semantically similar code diff paired with the commit message from the large parallel training corpus. The semantic similarity between two code diffs is measured by the cosine similarity of vectors obtained by a code diff encoder. Next, \Our treats the retrieved commit message as an example and uses it to guide the neural network to generate an understandable and concise commit message.

\subsection{Retrieval module}
In this module, we aim to retrieve the most semantically similar result. Specifically, we first train an encoder-decoder neural network on the large commit message generation dataset. The encoder is used to learn the semantics of code diffs and encode code diffs into a high-dimension semantic space. Then we retrieve the most semantically similar code diff paired with the commit message from the large parallel training corpus. The semantic similarity between two code diffs is measured by the cosine similarity of vectors obtained by a well-trained code diff encoder.

Recently, encoder-decoder neural network models ~\cite{LoyolaMM17,JiangAM17,jung2021commitbert}, which leverage an encoder to learn the semantic of code diff and employ a decoder to generate the commit message, have shown their superiority in the understanding of code diffs and commit messages generation. To enable the code diff encoder to understand the semantics of code diffs, we train it with a commit message generator on a large commit message generation dataset, which consists of about 0.9 million \lstinline{<code diff, commit message>} pairs.

To capture long-range dependencies (e.g. a variable is initialized before the changed line) and more contextual information of code diffs, we employ a Transformer-based encoder to learn the semantic representations of input code diffs. As shown in \Fig~\ref{fig:model}, a  Transformer-based encoder is stacked with multiple encoder layers. Each layer consists of four parts, namely, a multi-head self-attention module, a relative position embedding module, a feed forward network (FFN) and an add \& norm module. 
In $b$-th attention head, the input $\mathbf{X^{b}} = \left( \mathbf{x^{b}_1},\mathbf{x^{b}_2}, ... , \mathbf{x^{b}_l}\right )$  ( where $\mathbf{X^{b}} = \mathbf{X}[(b-1)*head_{dim}:b*head_{dim}]$,  $\mathbf{X}$ is the sequence of code diff embedding, $head_{dim}$ is the dimension of each head and $l$ is the input sequence length. ) is transformed to ($\mathbf{Head}^{b} = \mathbf{head_1^{b}},\mathbf{head_2^{b}}, ... , \mathbf{head_l^{b}}$) by:
\begin{equation}
\small
\begin{aligned}
\mathbf{head_i^{b}}&=\sum_{j=1}^{l} \alpha_{i j}\left( \mathbf{W_{V}x^{b}_j}+\mathbf{p_{i j}^{V}}\right)\\ 
e_{i j}&=\frac{(\mathbf{W_{Q}x^b_{i}}) ^{T}\left( \mathbf{W_{K}x^b_{j}}+\mathbf{p_{i j}^{K}}\right)}{\sqrt{d_{k}}}
\end{aligned}
\label{eq:self_attn}
\end{equation}
where $\alpha_{i j}=\frac{\exp e_{i j}}{\sum_{k=1}^{n} \exp e_{i k}}$, 
$\mathbf{W_Q}$, $\mathbf{W_K}$ and $\mathbf{W_V}$ are learnable matrix for queries, keys and values. $d_k$ is the dimension of queries and keys; $\mathbf{p_{i j}^{K}}$ and $\mathbf{p_{i j}^{V}}$ are relative positional representations for positions $i$ and $j$. 

The outputs of all heads are concatenated and then fed to the FFN modules which is a multi-layer perception. The add \& norm operation are employed after the multi-head attention and FFN modules. The calculations are as follows:
\begin{equation}
\small
\centering
\begin{aligned}
\mathbf{Head} &= Concat\left(\mathbf{Head^{1}, Head^{d}, Head^{B}} \right )\\
\mathbf{Hid} &= add\ \&\  norm \left(\mathbf{Head, X}\right )  \\
\mathbf{Enc} &= add\  \&\  norm \left(\mathbf{FFN\left(Hid\right ), Hid}\right ) \\
\end{aligned}
\label{eq:FFN}
\end{equation}
where $add\ \&\  norm \left( \mathbf{A_1,A_2} \right)  =  LN\left(\mathbf{A_1+ A_2}\right )$, $B$ is the number of heads and $LN$ is layer normalization. The  final output of encoder is sent to Transformer-based decoder to generate the commit message step by step. We use cross-entropy as loss function and adopt AdamW~\cite{loshchilov2017decoupled} to optimize the parameters of the code diff encoder and the decoder at the top of \Fig~\ref{fig:model}.

Next, the retrieval module is used to retrieve the most similar result from a large parallel training corpus. We firstly use the above code diff encoder to map code diffs into a high-dimensional latent space and retrieve the most similar example based on cosine similarity.

Specifically, after being trained in the commit message generation dataset, the code diff encoder can capture the semantic of code diff well. We use well-trained code diff encoder following a mean-pooling operation to map the code diff into a high dimensional space. Mathematically, given the input code diff embedding $\mathbf{X} = \left( \mathbf{x_1},\mathbf{x_2}, ... , \mathbf{x_l}\right )$, the code diff encoder can transformed them to $\mathbf{Enc} = \left( \mathbf{enc_1},\mathbf{enc_2}, ... , \mathbf{enc_l}\right )$. Then we obtain the semantic vector  of the code diff
by pooling operation:
\begin{equation}
\small
\label{eq:obain_vec}
\mathbf{vec} = pooling(\mathbf{Enc}) = mean \left( \mathbf{enc_1},\mathbf{enc_2}, ... , \mathbf{enc_l}\right ) \\
\end{equation}
\noindent where mean is a dimension-wise average operation.
We measure the similarity of two code diffs by cosine similarity of their semantic vectors and retrieve the most similar code diff paired with the commit message from the parallel training corpus. For each code diff, we return the first-ranked similar result. But, for the code diff in the training dataset, we return the second-ranked similar result because the first-ranked result is itself. 

\subsection{Generation module}
\label{sec:refine}

As shown at the bottom of \Fig~\ref{fig:model}, in the generation module, we treat the retrieved commit message as an exemplar and leverage it to guide the neural network model to generate an accurate commit message. Our generation module consists of three components: three encoders, an exemplar guider, and a decoder.

First, following \Eq~\ref{eq:self_attn},~\ref{eq:FFN}, three Transformer-based encoders are adopted to obtain the representations of the input 
code diff ($\mathbf{{Enc}^{d}} =\mathbf{enc}_1^{d},\mathbf{enc}_2^{d}, ... , \mathbf{enc}_l^{d}$), the similar code diff ($\mathbf{{Enc}^{s}} =\mathbf{enc}_1^{s},\mathbf{enc}_2^{s}, ... , \mathbf{enc}_m^{s}$), and similar commit message ($\mathbf{{Enc}^{m}} =\mathbf{enc}_1^{m},\mathbf{enc}_2^{m}, ... , \mathbf{enc}_n^{m}$) (step \textcircled{1} in \Fig~\ref{fig:model}), 
where subscripts $l,m,n$ are the length of the input code diff, the similar code diff, and the similar commit message, respectively. 

Second, since the retrieved similar commit messages may not always accurately describe the content/intent of the input code diffs even express totally wrong or irrelevant semantics. Therefore, we propose an \textit{exemplar guider} which first learns the semantic similarity between the retrieved and input code diff and then leverages the information of the similar commit messages based on the learned similarity to guide the commit message generation (step \textcircled{2} ). 
Mathematically, \textit{exemplar guider} calculate the semantic similarity ($\lambda$) between the input code diff and the similar code diff based on their representation $\mathbf{Enc}_l^{d}$ and $\mathbf{Enc}_m^{s}$ (step \textcircled{2} and \textcircled{3}):
\begin{equation}
\label{eq:sim}
\small
    \lambda = \sigma(\mathbf{W_s}[mean(\mathbf{Enc}^{d}),mean(\mathbf{Enc}^{s})]) 
\end{equation}
\noindent where $\sigma$ is the sigmoid activation function, $\mathbf{W_s}$ is a learnable matrix, and $mean$ is a dimension-wise average operation.

Third, we weight representations of code diff and similar commit message  by $1-\lambda$ and $\lambda$, respectively and then concatenate them to obtain the final input encoding.
\begin{equation}
\small
\mathbf{Enc^{dm}} = [(1-\lambda)*\mathbf{Enc^d} : \lambda*\mathbf{Enc^s} ]
\end{equation}

Finally, we use a Transformer-based decoder to generate the commit message. 
The decoder consists of multiply decoder layer and each layers includes a masked multi-head self-attention, a multi-head cross-attention module, a FFN module and an add \& norm module. Different from multi-head  self-attention module in the encoder, in terms of one token, masked multi-head self-attention in the decoder can only attend to the previous tokens rather than the before and after context. In $b$-th cross-attention layer, the input encoding ( $\mathbf{Enc^{dm}} = \left (\mathbf{enc^{dm}_1,enc^{dm}_2,...,enc^{dm}_{l+m} }\right)$) is queried by the output of the preceding commit message representations $\mathbf{Msg} = ( \mathbf{msg_1,...,msg_t})$ obtained by masked multi-head self-attention module.
\begin{equation}
\small
\begin{aligned}
Dec_{head_{i}^{b}}&=\sum_{j=1}^{l+m} \alpha_{i j} \left(\mathbf{W^{Dec}_{V} enc^b_{j}}\right)\\
Dec_{e_{i j}}&=\frac{\left(\mathbf{W^{Dec}_{Q}msg^b_{j}}) ^{T} (\mathbf{W^{Dec}_{K}enc^b_{i}}\right)}{\sqrt{d_{k}}}
\end{aligned}
\end{equation}
where $\alpha_{i j}=\frac{\exp Dec_{e_{i j}}}{\sum_{k=1}^{n} \exp Dec_{e_{i k}}}$, $\mathbf{W^{Dec}_{Q}}$, $\mathbf{W^{Dec}_{K}}$ and $\mathbf{W^{Dec}_{V}}$ are trainable projection matrices for queries, keys and values of the decoder layer. $\mathbf{t}$ is the length of preceding commit message. 

Next, we use \Eq~\ref{eq:FFN} to obtain the hidden states of each decoder layer. In the last decoder layers, we employ a MLP and softmax operator to obtain the generation probability of each commit message token on the vocabulary. Then we use the cross-entropy as the loss function and apply AdamW for optimization.

\section{Experimental Setup}

\subsection{Dataset}
\begin{table}[t]
\setlength{\tabcolsep}{5pt}
\small
\centering
\begin{tabular}{lccc} 
\toprule
Language & \text {Training} & \text {Validation} & \text {Test} \\
\midrule 
\text { \java } & 160,018 & 19,825&20,159 \\
\text { \csharp } &149,907 & 18,688 & 18,702 \\
\text { \cpp } & 160,948 & 20,000 &20,141 \\
\text { \python } & 206,777 &25,912 &25,837 \\
\text { \javascript } &197,529 &24,899 & 24,773 \\
\bottomrule
\end{tabular}
\label{tab:dataset}
\caption{Statistics of the evaluation dataset.}
\end{table}

In our experiment, we use a large-scale dataset \mcmd~\cite{tao2021evaluation} with five programming languages (PLs): \java, \csharp, \cpp, \python and \javascript. For each PL, \mcmd collects commits from the top-100 starred repositories on GitHub and then filters the redundant messages (such as rollback commits) and noisy messages defined in \citet{LiuXHLXW18}. Finally, to balance the size of data, they randomly sample and retain 450,000 commits for each PL. Each commit contains the code diff, the commit message, the name of the repository, and the timestamp of commit, etc.
To reduce the noise data in the dataset, we further filter out commits that contain multiple files or files that cannot be parsed (such as \texttt{.jar}, \texttt{.ddl}, \texttt{.mp3}, and \texttt{.apk}).

\subsection{Data pre-processing}

The code diff in \mcmd are based on line-level code change. To obtain more fine-grained code change, following previous study~\cite{PanthaplackelNG20}, we use a sequence of span of token-level change actions to represent the code diff. Each action is structured as \lstinline{<action> span of tokens <action end>}. There are four \lstinline{<action>} types, namely, \lstinline{<keep>}, \lstinline{<insert>}, \lstinline{<delete>}, and \lstinline{<replace>}. \lstinline{<keep>} means that the span of tokens are unchanged. \lstinline{<insert>} means that adding span of tokens. \lstinline{<delete>} means that deleting span of tokens. \lstinline{<replace>} means that the span of tokens in the old version that will be replaced with different span of tokens in the new version. Thus, we extend \lstinline{<replace>} to \lstinline{<replace old>} and \lstinline{<replace new>} to indicate the span of old and new tokens, respectively. 
We use difflib~\footnote{\url{https://docs.python.org/3/library/difflib.html}} to extract the sequence of code change actions.

\begin{table*}[t]
\scriptsize 
\setlength{\tabcolsep}{1.2pt}
\begin{tabular}{cl cccc cccc cccc cccc cccc} 
\toprule
 &\multirow{2}{*}{Model}   & \multicolumn{4}{c}{ \java}  & \multicolumn{4}{c}{ \csharp}  & \multicolumn{4}{c}{ \cpp}  & \multicolumn{4}{c}{ \python}   & \multicolumn{4}{c}{ \javascript}  \\ 
\cmidrule(r){3-6} \cmidrule(r){7-10} \cmidrule(r){11-14} \cmidrule(r){15-18} \cmidrule(r){19-22} 
 &&BLEU & Met. & Rou. &Cid.&BLEU & Met. & Rou. &Cid.&BLEU & Met. & Rou. &Cid.&BLEU & Met. & Rou. &Cid.&BLEU & Met. & Rou. &Cid. \\
  \midrule
 \multirow{2}{*}{IR-based} 
& \NNGen  &19.41 &12.40 &25.15 &1.23 &22.15 &14.77 &26.46 &1.55 &13.61 &9.39 &18.21 &0.73 &16.06 &10.91 &21.69 &0.92 &18.65 &12.50 &24.45 &1.21 \\
& \Lucene  &15.61 &10.56 &19.43 &0.94 &20.68 &13.34 &23.02 &1.36 &13.43 &8.81 &16.78 &0.67 &15.16 &9.63 &18.85 &0.85 &17.66 &11.25 &21.75 &1.02 \\
 \midrule
 \multirow{4}{*}{End-to-end} 
& \Commitgen  &14.07 &7.52 &18.78 &0.66 &13.38 &8.31 &17.44 &0.63 &11.52 &6.98 &16.75 &0.45 &11.02 &6.43 &16.64 &0.42 &18.67 &11.88 &24.10 &1.08 \\
& \Codisum  &13.97 &6.02 &16.12 &0.39 &12.71 &5.56 &14.40 &0.36 &12.44 &6.00 &14.39 &0.42 &14.61 &8.59 &17.02 &0.42 &11.22 &5.32 &13.26 &0.28 \\
& \NMT  &15.52 &8.91 &21.13 &0.86 &12.71 &8.11 &17.16 &0.62 &11.57 &7.06 &17.46 &0.51 &11.41 &7.18 &18.43 &0.48 &18.22 &12.07 &24.43 &1.12 \\
& \Ptrnet  &17.71 &11.33 &24.32 &0.99 &15.98 &10.18 &21.16 &0.83 &14.06 &9.63 &20.17 &0.63 &15.89 &11.36 &23.49 &0.76 &20.78 &14.52 &27.87 &1.29 \\

\midrule
 \multirow{2}{*}{Hybrid} 
 &\ATOM &16.42 &11.66 &22.67 &0.91 &/ &/ &/ &/ &/ &/ &/ &/ &/ &/ &/ &/ &/ &/ &/ &/ \\
& \Corec  &18.51 &11.26 &24.78 &1.13 &18.41 &11.70 &23.73 &1.12 &14.02 &8.63 &20.10 &0.72 &15.09 &9.60 &22.35 &0.80 &21.30 &13.84 &27.53 &1.40 \\
\midrule
 \multirow{3}{*}{Pre-trained} 
& \commitbert  &22.32 &12.63 &28.03 &1.42 &20.67 &12.31 &25.76 &1.25 &16.16 &10.05 &19.90 &0.94 &17.29 &11.31 &22.36 &1.01 &23.40 &15.64 &30.51 &1.54 \\
& \codetf-small  &22.28 &14.16 &29.71 &1.37 &18.92 &11.71 &24.95 &1.05 &16.08 &11.19 &21.60 &0.79 &17.49 &12.46 &24.65 &0.90 &21.97 &14.48 &28.65 &1.42 \\
& \codetf-base  &22.76 &14.57 &30.23 &1.43 &22.21 &14.51 &29.08 &1.33 &16.73 &11.69 &22.86 &0.85 &17.99 &12.74 &25.27 &0.96 &22.87 &15.12 &29.81 &1.50 \\
\midrule
 \multirow{2}{*}{\textbf{Ours} } & \multirow{2}{*}{\textbf{\Our}} 
 &\textbf{25.66} &\textbf{15.46} &\textbf{32.02} &\textbf{1.76} &\textbf{26.33} &\textbf{16.37} &\textbf{31.31} &\textbf{1.84} &\textbf{19.13} &\textbf{12.55} &\textbf{24.52} &\textbf{1.14} &\textbf{21.79} &\textbf{14.68} &\textbf{28.35} &\textbf{1.40} &\textbf{25.55} &\textbf{16.31} &\textbf{31.79} &\textbf{1.84} \\
&&$\uparrow$\textbf{13\%} &$\uparrow$\textbf{6\%} &$\uparrow$\textbf{6\%} &$\uparrow$\textbf{23\%} &$\uparrow$\textbf{19\%} &$\uparrow$\textbf{13\%} &$\uparrow$\textbf{8\%} &$\uparrow$\textbf{38\%} &$\uparrow$\textbf{14\%} &$\uparrow$\textbf{7\%} &$\uparrow$\textbf{7\%} &$\uparrow$\textbf{34\%} &$\uparrow$\textbf{21\%} &$\uparrow$\textbf{15\%} &$\uparrow$\textbf{12\%} &$\uparrow$\textbf{46\%} &$\uparrow$\textbf{12\%} &$\uparrow$\textbf{8\%} &$\uparrow$\textbf{7\%} &$\uparrow$\textbf{23\%}\\
\midrule 
Ablation & \Our\text{-}Guider &23.37 &13.98 &30.01 &1.53 &21.33 &13.56 &27.33 &1.31 &17.43 &12.10 &22.03 &0.95 &19.44 &13.89 &26.4 &1.01 &23.39 &15.64 &30.51 &1.54 \\
\bottomrule
\end{tabular}
\caption{Comparison of \Our with baselines under four metrics on five programming languages. Met., Rou., and Cide. are short for Meteor, Rouge-L, and Cider, respectively. All results are statistically significant (with $p <0.01$).}
\label{tab:compare_with_baselines}
\end{table*}

\subsection{Hyperparameters} 
We follow~\cite{tao2021evaluation} to set the maximum lengths of \diff and commit message to 200 and 50, respectively. We use the weight of the encoder of \codetf-base~\cite{0034WJH21} to initialize the code diff encoders and use the decoder of \codetf-base to initialize the decoder in \Fig~\ref{fig:model}. 
The original vocabulary sizes of \codetf is 32,100. We add nine special tokens (\lstinline{<keep>}, \lstinline{<keep_end>}, \lstinline{<insert>}, \lstinline{<insert_end>}, \lstinline{<delete>}, \lstinline{<delete_end>}, \lstinline{<replace_old>}, \lstinline{<replace_new>}, and \lstinline{<replace_end>}) and the vocabulary sizes of code and queries become 32109.
For the optimizer, we use AdamW with the learning rate 2e-5. The batch size is 32. The max epoch is 20. In addition, we run the experiments 3 times with random seeds 0,1,2 and display the mean value in the paper. The experiments are conducted on a server with 4 GPUs of NVIDIA Tesla V100 and it takes about 1.2 hours each epoch.

\subsection{Evaluation metrics} 

We evaluate the quality of the generated messages using four metrics: BLEU~\cite{PapineniRWZ02}, Meteor~\cite{BanerjeeL05}, Rouge-L~\cite{lin-2004-rouge}, and Cider~\cite{VedantamZP15}. These metrics are prevalent metrics in machine translation, text summarization, and image captioning. There are many variants of BLEU being used to measure the generated message, We choose B-Norm (the BLEU result in this paper is B-Norm), which correlates with human perception the most~\cite{tao2021evaluation}.
The detailed metrics calculation can be found in Appendix.

\subsection{Baselines}
We compare \Our{} with four end-to-end neural-based models, two IR-based methods, two hybrid approaches which combine IR-based techniques and end-to-end neural-based methods, and three pre-trained-based models. Four end-to-end neural-based models include
\Commitgen~\cite{JiangAM17}, \Codisum~\cite{Xu00GT019}, \NMT~\cite{LoyolaMM17}, \Ptrnet~\cite{LiuLZFDQ19} and \ATOM~\cite{Liu20Atom}. They all train models from scratch.
Two IR-based methods are \NNGen~\cite{LiuXHLXW18} and \Lucene~\cite{lucene}, they retrieve the similar code diff based on different similarity measurements and reuse the commit message of the similar code diff as the final result. \Corec and \ATOM are all hybrid models which combine the neural-based models and IR-based techniques. 
Three pre-trained models are \commitbert, \codetf-small, and \codetf-base. They are pre-trained on the large parallel code and natural language corpus and fine- tuned on the commit message generation dataset. All baselines except \Lucene, \codetf-small and \codetf-base are introduced in \Sec ~\ref{sec:related_work}. \Lucene is a traditional IR baseline, which uses TF-IDF to represent a code diff as a vector and searches the similar \diff based on the cosine similarity between two vectors. \codetf-small and \codetf-base are source code pre-trained models and have achieved promising results in many code-related tasks~\cite{0034WJH21}. We fine-tune them on \mcmd as strong baselines. 
In addition, we only evaluate \ATOM on \java dataset as the current implementation of \ATOM only supports Java.

\section{Experimental Results}
\subsection{How does \Our{} perform compared with baseline approaches?}

To evaluate the effectiveness of \Our{}, we conduct the experiment by comparing it with the 11 baselines including two IR-based approaches, four end-to-end neural-based approaches, two hybrid approaches, and three pre-train-based approaches in terms of four evaluation metrics. The experimental results are shown in \Tab~\ref{tab:compare_with_baselines}.

We can see that IR-based models \NNGen and \Lucene generally outperform end-to-end neural models on average in terms of four metrics. It indicates that retrieved similar results can provide important information for commit message generation. \Corec, which combines the IR-based method and neural method, performs better than \NNGen on \cpp and \javascript dataset but lower than \NNGen on \java, \csharp and \python. This is because \Corec only leverages the information similar code diff at the inference stage. \ATOM, which priorities the generated result of the neural-based model and retrieved result of the IR-based method, also outperforms the IR-based approach \Lucene and three neural-based models \Commitgen, \Codisum, and \NMT.
Three pre-trained-based approaches outperform other baselines in terms of four metrics on average. 
\codetf-base performs best among them on average. Our approach performs the best among all approaches on 5 programming languages in terms of four metrics. This is because \Our{} treats the retrieved similar commit message as an exemplar and leverages it to guide the neural network model to generate an accurate commit message.

\begin{figure}[ht]
\centering
\small
\renewcommand\arraystretch{1.1}
\scalebox{0.87}{
\begin{tabular}{p{0.238\linewidth}l}
\toprule
\colorbox{lime}{Code Diff} \\
\multicolumn{2}{l}{\includegraphics[width=1.0\linewidth]{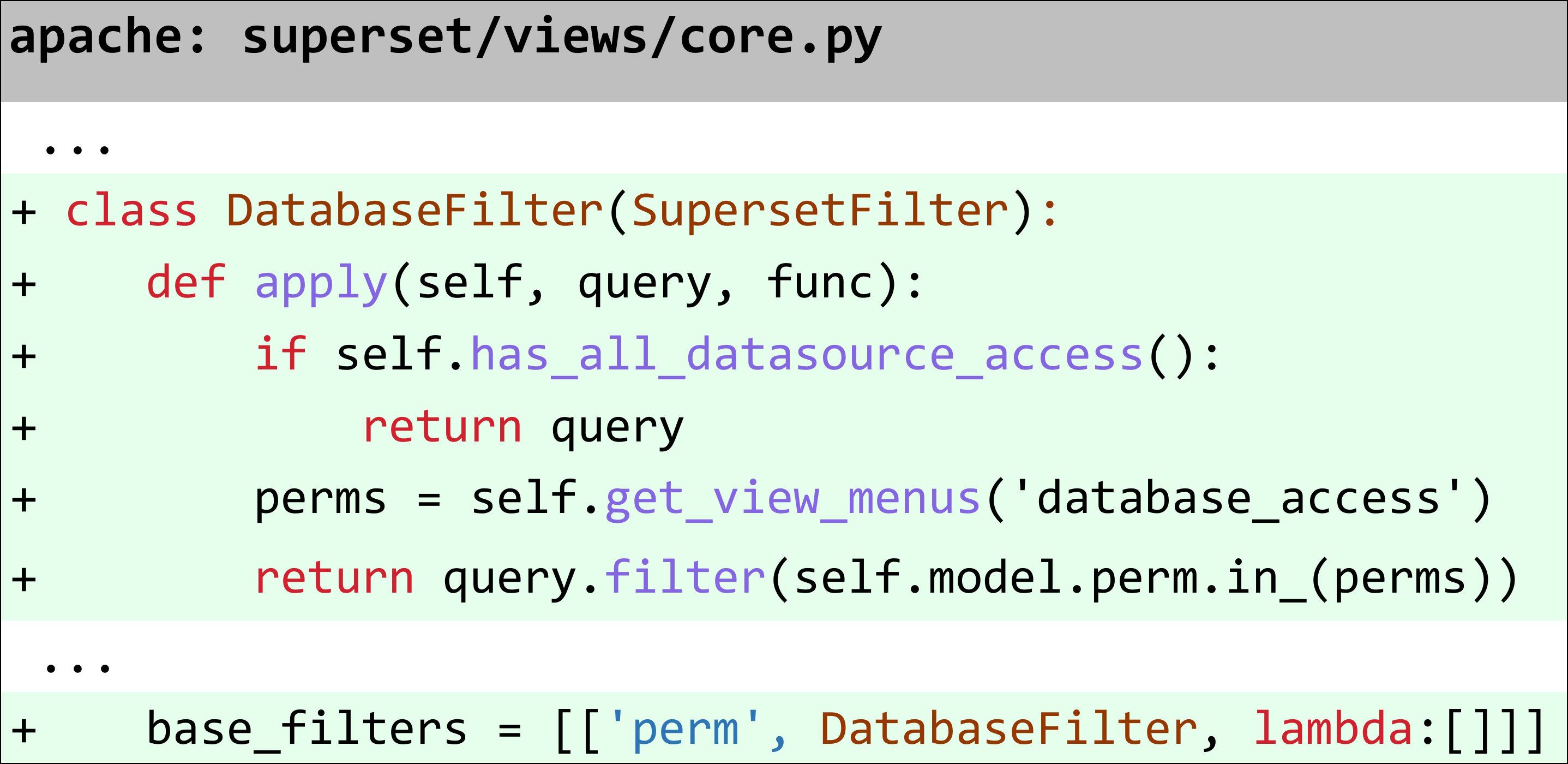}} \\
 \colorbox{light-gray}{Reference} & \colorbox{light-gray}{\emph{Filter out unavailable databases }}\\ 
 \midrule
 \textbf{Baselines} \\
\NNGen & Revert `` Filter out unavailable databases''\\ 
\Lucene &Revert `` filter out unavailable databases ''  \\ 
\Commitgen  &Merge pull request from mistercrunch / UNK \\ 
\NMT  &Add <unk> to <unk>\\ 
\Ptrnet  &Add support for dashboards in database\\
\Corec  &Remove <unk>\\ 
\commitbert  &Add DatabaseFilter ( )\\
\codetf-small  &[database] Add databasefilter to filter all users\\
\codetf-base &[hotfix] Adding databasefilter to core.py  \\
\midrule
 \multirow{2}{*}{\textbf{\Our{}}}  &Stage \uppercase\expandafter{\romannumeral1} : Revert `` Filter out unavailable databases '' \\  
 &Stage \uppercase\expandafter{\romannumeral2} : \textbf{Filter out unavailable databases} \\ 
\bottomrule
\end{tabular}
}
\caption{An example of generated commit messages. Reference is the developer-written commit message.  The results of our approach in stage \uppercase\expandafter{\romannumeral1} and \uppercase\expandafter{\romannumeral2} are returned by the retrieved module and generation module, respectively.}
\label{tab:case_1}
\end{figure}

We also give an example of commit messages generated by our approach and the baselines in Figure~\ref{tab:case_1}.
IR-based methods \NNGen and \Lucene can retrieve semantically similar but not completely correct commit message. Specifically, retrieved commit messages contain not only the important semantic (``Filter out unavailable databases'') of the current code diff but also the extra information (``Revert''). Neural network models generally capture the action of ``add'' but fail to further understand the intend of the code diff. The hybrid model \Corec cannot generate the correct commit message either. Our model treats the retrieved result (Revert "Filter out unavailable databases'') as an exemplar, and guides the neural network model to generate the correct commit message.

\subsection{What is the effectiveness of exemplar guider? }

\revised{We conduct the ablation study to verify the effectiveness of \textit{exemplar guider} module. Specifically, as shown at the bottom of~\Fig~\ref{fig:model}, we directly concatenated the representations of retrieved results and fed them to the decoder to generate commit messages without using the \textit{exemplar guider}. As shown at the bottom of the~\Tab~\ref{tab:compare_with_baselines}, we can see that the performance of the ablated model (\Our\text{-}Guide) degrades in all programming languages in terms of four metrics. It demonstrates the effectiveness of our \textit{exemplar guider}.}

\subsection{What is the performance when we reteieve $k$ relevant commits?}
\begin{figure}[ht]
    \includegraphics[width=1.01\linewidth]{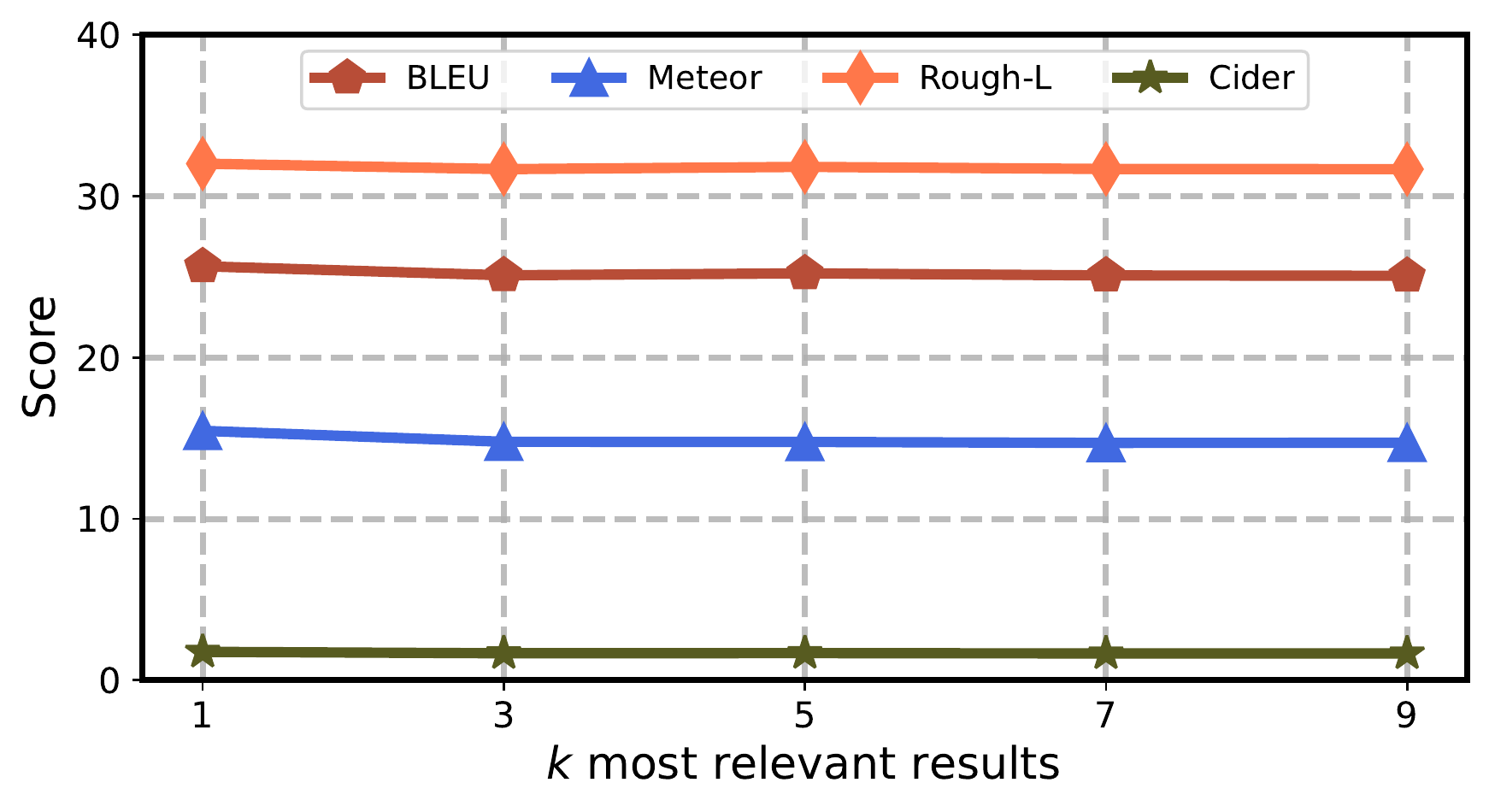}
    \caption{Performance of models augemented with $k$ retrieved relevant commits.}
    \label{fig:diff_k}
\end{figure}

\revised{We also conduct experiments to recall $k$ ($k$=1, 3, 5, 7, 9) most relevant commits to augment the generation model. Specifically, as shown in \Fig~\ref{fig:model} the relevance of the code diffs is measured by the cosine similarity their semantic vectors obtained by \Eq~\ref{eq:obain_vec}. Then retrieved $k$ relevant commits are encoded and fed to the exemplar guidar to obtain semantic similarities by \Eq~\ref{eq:sim}, respectively. Finally, we weight representations of code diff and similar commit messages according to the semantic similarities and feed them to the decoder to generate commit messages step by step. The experimental results are shown in \Fig~\ref{fig:diff_k}. We can see that the performance is generally stable on different $k$. In our future work, we will continue to study alternatives on leveraging the information of the retrieved results, e.g., how many commits to retrieve and how to model the corresponding information.}

\subsection{Can our framework boost the performance of existing models?}
\label{sec:RQ_adaption}
\begin{figure}[ht]
    \includegraphics[width=1.0\linewidth]{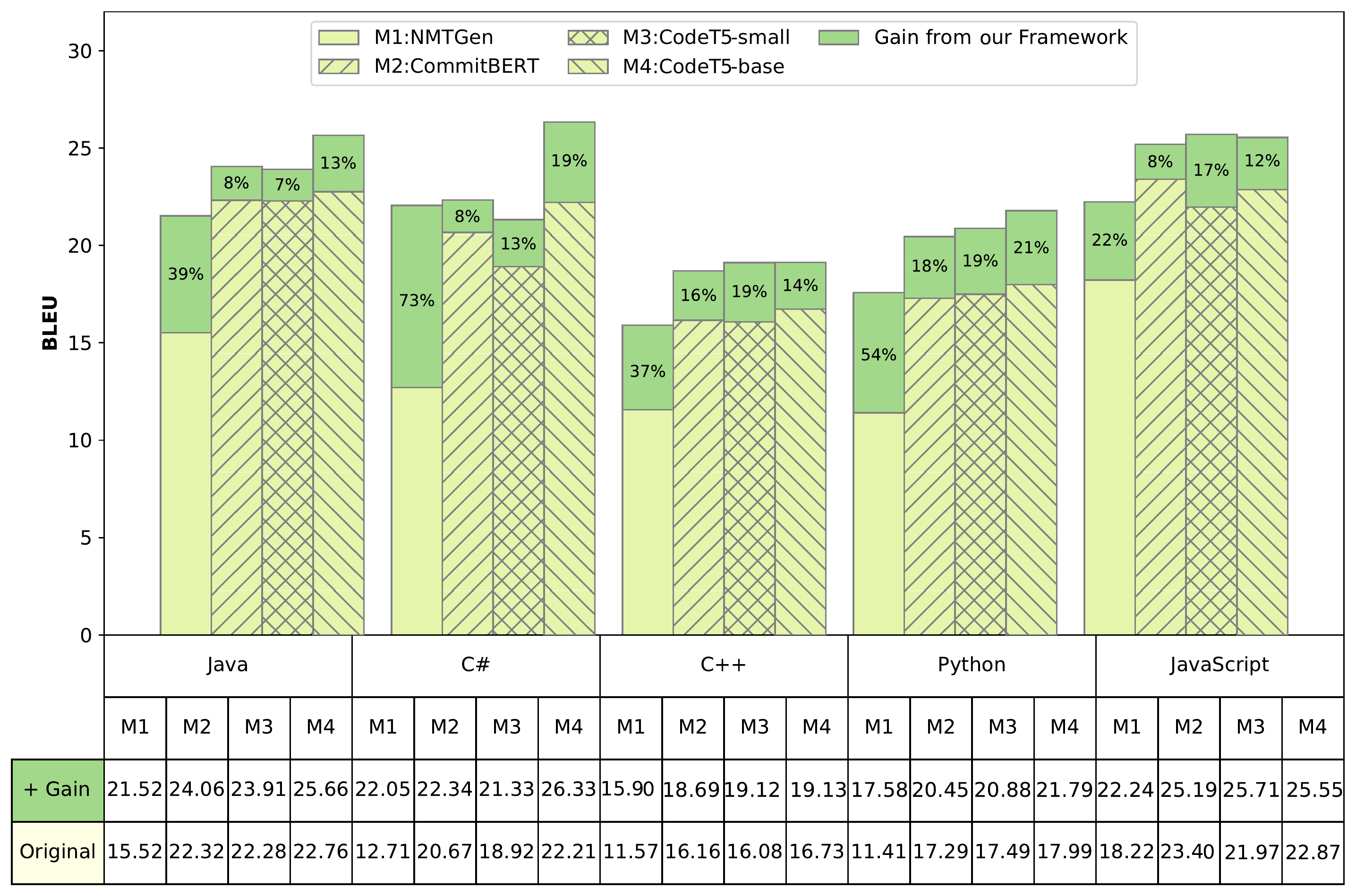}
    \caption{Performance gains on four models. The original performance of the models are in yellow and gains from our framework are in green. The percentage value in each bar is the rate of improvement.}
    \label{fig:gain}
\end{figure}

We further study whether our framework can enhance the performance of the existing Seq2Seq neural network model in commit message generation. Therefore,  we adapt our framework to four Seq2Seq-based models, namely \NMT (\Mone), \commitbert (\Mtwo), \codetf-small (\Mthree) and \codetf-base (\Mfour). Specifically, we use the encoder of these models as our code diff encoder and obtain the high-dimensional semantic vectors in the retrieval module (\Fig~\ref{fig:model}). In the generation module, we use the encoder of their models to encode input code diffs, similar code diffs, and similar commit messages. We also use the decoder of their models to generate commit messages.

The experimental results are shown in \Fig~\ref{fig:gain}, we present the performance of four original models (yellow) and gains (green) from our framework on five programming languages in terms of BLEU~\footnote{We show results of other three metrics in Appendix due to space limitation. Our conclusions also hold.}  score. Overall, we can see that our framework can improve the performance of all four neural models in all programming languages. Our framework can improve the performance of the original model from 7\% to 73\%.
Especially, after applying our framework, the performance of \NMT has more than 20\% improvement on all programming languages. In addition, Our framework can \textit{boost} the performance of 
\NMT on BLUE, Meteor, Rouge-L, and 
Cider by 43\%, 49\%, 33\%, and 61\% on average,  \textit{boost}
\commitbert by 11\%, 9\%, 11\%, and 12\%, \textit{boost} 
\codetf-small by 15\%, 14\%, 11\%, and 26\%, and \textit{boost} 
\codetf-base by 16\%, 10\%, 8\%, and 32\%~\footnote{The result can be found in 1\text{-}4 of Appendix}.

\subsection{Human evaluation}

\begin{table}[t]
\centering 
\small
\setlength{\tabcolsep}{1.3pt}
\begin{tabular}{lccc}
\toprule
Model       &Informativeness        &Conciseness            &Expressiveness \\ 
\midrule
\commitbert{}     &1.22 ($\pm$1.02)  &2.03 ($\pm$1.04)   &2.46 ($\pm$0.99)  \\
\NNGen{}          &1.03 ($\pm$1.00)   &1.74 ($\pm$1.01)   &2.36 ($\pm$0.95)  \\
\NMT{}            &0.74 ($\pm$0.92)  &1.56 ($\pm$0.93)   &2.11 ($\pm$0.94)  \\
\Corec{}          &1.05 ($\pm$1.09)  &1.80 ($\pm$1.05)    &2.43 ($\pm$0.88)  \\
\midrule
\Our{}            &\textbf{2.49} ($\pm$1.10)   &\textbf{3.08} ($\pm$0.96)   &\textbf{2.85} ($\pm$0.84)  \\
\bottomrule
\end{tabular}
\caption{Results of human evaluation (standard deviation in parentheses).} 
\label{tab:human_evaluation}

\end{table}

We also conduct a human evaluation by following the previous works~\cite{MorenoASMPV13,PanichellaPBZG16,Shi0D0HZS21} to evaluate the semantic similarity of the commit message generated by \Our{} and four baselines  \NNGen{}, \NMT, \commitbert{}, and \Corec. The four baselines are IR-based, end-to-end neural network-based, hybrid, and pre-trained-based approaches, respectively. We randomly choose 50 code diff from the testing sets and their commit message generated by four approaches. Finally, we sample 250 \lstinline{<code diff, commit message>} pairs to score. Specifically, we invite 4 volunteers with excellent English ability and more than three years of software development experience.  Each volunteer is asked to assign scores from 0 to 4 (the higher the better) to the generated commit message from the three aspects: \textbf{Informativeness} (the amount of important information about the code diff reflected in the commit message),  \textbf{Conciseness} (the extend of extraneous information included in the commit message), and \textbf{Expressiveness} (grammaticality and fluency). Each pair is evaluated by four volunteers, and the final score is the average of them.

To verify the agreement among the volunteers, we calculate the Krippendorff’s alpha~\cite{hayes2007answering}  and Kendall rank correlation coefficient (Kendall’s Tau) values~\cite{kendall1945treatment}. The value of Krippendorff’s alpha is 0.90 and the values of pairwise Kendall’s Tau range from 0.73 to 0.95, which indicates that there is a high degree of agreement between the 4 volunteers and that scores are reliable. \Tab~\ref{tab:human_evaluation} shows the result of human evaluation. \Our{} is better than other approaches in Informative, Conciseness, and Expressiveness, which means that our approach tends to generate concise and readable commit messages with more comprehensive semantics. In addition, we confirm the superiority of our approach using Wilcoxon signed-rank tests~\cite{wilcoxon1970critical} for the human evaluation. Results~\footnote{Available in Appendix} show that the improvement of \Our{} over other approaches is statistically significant with all p-values smaller than 0.05 at 95\% confidence level.

\section{Conclusion}

This paper proposes a new retrieval-augmented neural commit message generation method, which treats the retrieved similar commit message as an exemplar and uses it to guide the neural network model to generate an accurate and readable commit message. Extensive experimental results demonstrate that our approach outperforms recent baselines and our framework can significantly boost the performance of four neural network models. Our data, source code and Appendix are available at \url{https://github.com/DeepSoftwareAnalytics/RACE}. 

\section*{Limitations}

We have identified the following main limitations:

\emph{Programming Languages.} We only conduct experiments on five programming languages. Although in principle, our framework is not specifically designed for certain languages, models perform differently in different programming languages. Therefore, more experiments are needed to confirm the generality of our framework. In the future, we will extend our study to other programming languages.

\emph{Code base.} Compared with purely neural network-based models, our method needs a code base to retrieve the most similar example from that. This limitation is inherited from IR-based techniques.

\emph{Training Time.} In addition to modeling the information of input code diffs, our model needs to retrieve similar diffs and encode them. Thus, our model takes a long time to train (about 35 hours to train the model).

\emph{Long Code Diffs}. Longer code diffs may contain more complex semantics or behaviors.  
Long diffs (over 512 tokens) are truncated in our approach and some information would be lost. In our future work, we will design mechanisms to better handle long diffs.

\section*{Acknowledgement}
We thank reviewers for their valuable comments on this work. This research was supported by National Key R\&D Program of China (No. 2017YFA0700800).
We would like to thank Jiaqi Guo and Wenchao Gu for their valuable suggestions and feedback during the work discussion process. We also thank the participants of our human evaluation for their time. 

\bibliographystyle{acl_natbib}
\bibliography{ref}

\end{document}

%% file: tool.tex
\usepackage{enumitem}
\usepackage{makecell}
\usepackage{xspace}
\usepackage{nicematrix}
\usepackage{amsmath,amsfonts}

\usepackage{amsmath,amssymb,amsfonts}
\usepackage{graphicx}
\usepackage{textcomp}
\usepackage{subfigure}
\usepackage{tcolorbox}
\usepackage{booktabs}
\usepackage{tabularx}
\usepackage{lipsum}
\usepackage{multirow}

\newcommand{\myauthornote}[3]{}

\newcommand{\revised}[1]{{#1}}
\newcommand{\later}[1]{}

\newcommand{\diff}{code diff\xspace}

\newcommand{\java}{Java\xspace}
\newcommand{\csharp}{C\#\xspace}
\newcommand{\cpp}{C++\xspace}
\newcommand{\python}{Python\xspace}
\newcommand{\javascript}{JavaScript\xspace}

\newcommand{\Commitgen}{CommitGen\xspace}
\newcommand{\NMT}{NMTGen\xspace}
\newcommand{\Codisum}{CoDiSum\xspace}
\newcommand{\NNGen}{NNGen\xspace}
\newcommand{\ATOM}{ATOM\xspace}

\newcommand{\Ptrnet}{PtrGNCMsg\xspace} 
\newcommand{\Corec}{CoRec\xspace}
\newcommand{\Lucene}{Lucene\xspace}

\newcommand{\commitbert}{CommitBERT\xspace}
\newcommand{\codetf}{CodeT5\xspace}
\newcommand{\Our}{RACE\xspace} 

\newcommand{\mcmd}{MCMD\xspace} 

\newcommand{\Mone}{M1\xspace}
\newcommand{\Mtwo}{M2\xspace}
\newcommand{\Mthree}{M3\xspace}
\newcommand{\Mfour}{M4\xspace}

\newcommand{\Fig}{Figure\xspace}
\newcommand{\Tab}{Table\xspace}
\newcommand{\Sec}{Section\xspace}
\newcommand{\Eq}{Equation\xspace}

\usepackage{xcolor}
\definecolor{light-gray}{gray}{0.85}
\definecolor{light-blue}{RGB}{177,206,70} 